\documentclass[draft]{agujournal2019}
\usepackage{url} 
\usepackage{lineno}
\usepackage[inline]{trackchanges} 
\usepackage{soul}
\usepackage[utf8x]{inputenc} 
\draftfalse

\journalname{Radio Science}

\begin{document}

%
%

\title{Ionospheric response to Strong Geomagnetic Storms during 2000-2005: An IMF clock angle perspective}

%
%




\authors{Sumanjit Chakraborty\affil{1}, Sarbani Ray\affil{3}, Abhirup Datta\affil{1,2}, Ashik Paul\affil{3}}


\affiliation{1}{Discipline of Astronomy, Astrophysics and Space Engineering, IIT Indore, Simrol, Indore 453552, Madhya Pradesh, India}
\affiliation{2}{Center for Astrophysics and Space Astronomy, Department of Astrophysical and Planetary Science, University of Colorado, Boulder, CO 80309, USA}
\affiliation{3}{Institute of Radio Physics and Electronics, University of Calcutta, Kolkata 700 009, West Bengal, India}
\vspace{20pt} \hspace{80pt}
\textbf{\textit{Accepted for publication in Radio Science}}




\correspondingauthor{Sumanjit Chakraborty}{sumanjit11@gmail.com}




\begin{keypoints}
\item Strong geomagnetic storms during 2000–2005 are selected in terms of Dst and IMF $B_z$.
\item Global ion density plots are used to study the effects of PPEF. 
\item Longitudes of ESF occurrence in response to PPEF are predicted from the time of northward to southward transition of the IMF clock angle.
\end{keypoints}

%
%

%
%


\begin{abstract}
This paper presents the equatorial ionospheric response to eleven strong-to-severe geomagnetic storms that occurred during the period 2000-2005, the declining phase of the solar cycle 23. The analysis has been performed using the global ion density plots of Defense Meteorological Satellite Program (DMSP). Observations show that for about $91\%$ of the cases, post-sunset equatorial irregularities occurred within 3h from the time of northward to southward transition of the Interplanetary Magnetic Field (IMF) clock angle, thus bringing out the importance of the role played by IMF $B_y$ in the process of Prompt Penetration of Electric Field (PPEF) in addition to the IMF $B_z$. This is an improvement from the previously reported \cite{sc:7} 4h window of ESF generation from the southward IMF $B_z$ crossing -10 nT. 
\end{abstract}

%
%

\section{Introduction}
The solar drivers of geomagnetic storms are the Coronal Mass Ejection (CME) and the Co-rotating Interaction Region (CIR). CMEs pass across the Earth frequently with an average rate of two events per month, having variations throughout a given solar cycle \cite{sc:13}. The occurrence of CME-driven geomagnetic storms having stronger intensity tends to be high during the ascending and solar maximum phases in a solar cycle.
The other type, CIRs, are the drivers of weak-to-moderate geomagnetic storms that have a similar impact on the ionosphere comparable to a CME driven storm, and they mostly occur during the declining phase of a solar cycle \cite{sc:23,sc:24,sc:14,sc:22}. When the high speed solar wind streams, originating from a coronal hole, interact with the slow solar winds, shocks along with compression and rarefaction regions are formed, causing recurrent geomagnetic storm effects on the Earth \cite{sc:15,sc:16}.

The solar wind emanating from the Sun carries a frozen-in Interplanetary Magnetic Field (IMF) that interacts with the Earth's magnetosphere-ionosphere system. The IMF B field defined as: 
\begin{equation}
B = \sqrt{B_y{^2}+B_z{^2}}
\end{equation}
is a 2D vector (the y and z components designated as $B_y$ and $B_z$ respectively) of the solar wind IMF. The vertical plane, with respect to the ecliptic, is the $B_y-B_z$ plane with $B_x$ being the corresponding x component (Sun-to-Earth line) of solar wind IMF. When $B_z$ is southward or has a negative orientation, the coupling to geomagnetic field is the strongest, setting conditions favorable for geomagnetic storm activity. While IMF B gives the magnitude, the orientation of this field is given by the IMF clock angle ($\theta$) which is the angle produced in the vertical plane from the vector addition of the $B_y$ and $B_z$ components of IMF and is defined as:
\begin{equation}
tan\theta = \frac{B_y}{B_z}    
\end{equation}
where $-180^{\circ}\leq\theta\leq180^{\circ}$ \cite{sc:25}.
The Interplanetary Electric Field of the solar wind enters the Earth's ionosphere via magnetosphere under geomagnetically perturbed conditions due to the southward turning of the Interplanetary Magnetic Field's (IMF $B_z$) north-south component \cite{sc:17} passing across Earth for a long time interval \cite{sc:3}. A sudden increase is observed in the dawn-to-dusk polar cap potential, which results from the change in region 1 current due to the passage of IMF $B_z$.  An undershielding condition gets developed when the ionosphere adapts itself from this prompt electric field from the outer magnetosphere. This happens because the region 2 current, that shields the low-latitude ionosphere from electric fields at high latitudes, varies slowly in comparison to the region 1 current. As a result, entry of this electric field from the high latitudes to the equatorial latitudes occur promptly and hence are known as the Prompt Penetration Electric Field (PPEF) \cite{sc:2}. As time passes, an inertial field gets developed in the inner magnetosphere, that opposes the PPEF and produces shielding. When there is a sudden northward turn of the IMF $B_z$ that cancels this penetrating electric field, the inertial field, which is oppositely directed to the incoming electric field becomes dominant in the ionosphere causing an overshielding condition \cite{sc:18}. The electric field enhancements in the low-latitude ionosphere are related to magnetic activity and occur during the main phase of magnetic storms, revealing the fact that the interplanetary electric field continuously penetrates to the low-latitude ionosphere without shielding for many hours as long as the strengthening of the magnetic activity is going on under storm conditions \cite{sc:35}.

In addition to PPEF, the electric field, which influences low-to-equatorial transportation of plasma during the disturbed conditions, comes from the neutral wind circulation changes at the sub-auroral thermosphere as a result of the deposition of enormous energy from the solar wind-magnetosphere and ionosphere coupling. This is known as the Disturbance Dynamo Electric Field (DDEF), which opposes the PPEF and lasts upto few days after becoming active a few hours post-PPEF \cite{sc:19}. 
Furthermore, the PPEF is directed eastward till 22:00 Local Time (LT), and turns westward after that and remains so till morning hours \cite{sc:20}. During the daytime, eastward directed dynamo electric field of the E region gives rise to the E $\times$ B drift (where B is the magnetic fields that are nearly parallel to the Earth's surface at equatorial latitudes) in the vertical direction causing an upward lift of plasma in the F region at the magnetic equator. As a consequence of forces due to pressure-gradient and gravity, these plasma move along the magnetic field lines. Thus the Equatorial Ionization Anomaly (EIA) is formed that has two crests around $\pm$15$^{\circ}$ magnetic latitudes and trough around the magnetic equator \cite{sc:10}. 

Ionosphere over the equatorial and low latitude regions present a dynamic feature, in addition to the EIA, known as the Equatorial Spread F (ESF) or irregularities of plasma that are plasma structures having scale sizes ranging from a few meters to a few hundred kilometers \cite{sc:30,sc:31,sc:29,sc:6} and references therein. The dynamo electric fields and plasma densities decrease in the E region around local sunset and weaken the EIA. However, simultaneously the F layer dynamo is intensified. The ionospheric plasma is transported upward by the post-sunset electric field, which enhances the anomaly crests. The eastward PPEF gets enhanced around local sunset because of the gradient in day to night conductivity. The enhanced electric field generates Rayleigh-Taylor instability of plasma. The instability causes the formation of irregularities, which are large, plasma depleted structures known as the Equatorial Plasma Bubbles (EPB) \cite{sc:30,sc:29,sc:11} and references therein. These EPBs cause satellite signals to scintillate. In recent years, the study of PPEF has become a vital space weather issue as they are related to these scintillations that cause catastrophic effects on the various Global Navigation Satellite Systems (GNSS) such as GPS, GLONASS, etc. and Regional Navigation Satellite Systems (RNSS) such as NavIC/IRNSS. 
Studies have been performed by several researchers with the Defense Meteorological Satellite Program (DMSP) satellite in situ measurements. \citeA{sc:2} studied large geomagnetic storms of solar cycle 23 and showed that with the knowledge of the time duration of the main phase of storms, one would be able to determine the dusk sector corresponding to the main phase and would be able to specify the longitude interval over which the scintillations could be detected. \citeA{sc:7} have showed that in a longitude sector where the local time is dusk, ESF gets generated within 4 hours of the southward turning of IMF $B_z$.

The equatorial ionospheric response to PPEF for the strong (G3 class, K$_p$ = 7) and severe (G4 class, K$_p$ = 8), according to the National Oceanic and Atmospheric Administration (NOAA)  space weather scales, geomagnetic storms during the period from 2000-2005, which fall in the declining phase of solar cycle 23, using in situ DMSP global ion density measurements, have been studied in this work. This paper presents the ionospheric response to strong-to-severe geomagnetic storms during the declining phase of solar cycle 23 from an IMF clock angle perspective.

\section{Data}

The storms in this paper are selected on the basis of the Disturbance storm time (Dst) index (nT) obtained from http://wdc.kugi.kyoto-u.ac.jp/dstdir/. The 1 minute high resolution interplanetary parameters: $B_y$ (nT) and $B_z$ (nT) component of the IMF respectively, along with the SYM-H (nT) index are obtained from https://omniweb.gsfc.nasa.gov. The in situ total ion density measurements are obtained from https://cindispace.utdallas.edu/DMSP by the DMSP f12, f14, and f15 sun-synchronous, near-polar orbiting satellites, sampling with a time interval of 4s at 840 km altitude with an orbital period of 101 minutes and crossing the magnetic equator during the time span of 19:00-22:00 Magnetic Local Time (MLT).

\section{Results and Discussions}

During the period from 2000-2005, eleven storms have been selected that satisfied the criteria of qualifying for a strong storm i.e., Dst $\leq$ -100 nT and IMF $B_z$ $\leq$ -10 nT for at least 3 hours \cite{sc:21,sc:3}. Table 1 shows a summary of the storm particulars (along with the time and day of Dst minimum), wherein two storms (April 06, 2000 and August 12, 2000) showed minimum Dst values below -200 nT qualifying them to be falling under the severe storm category, rest falling under the strong storm category \cite{sc:12}. The storms of April 06, 2000 and May 30, 2005 have been discussed in this paper. 
\begin{table*}
\begin{center}
\begin{tabular}{|l|c|c|c|c|}
\hline 
Period & Minimum Dst(nT) & UT(HH:MM) & DOY(DD) \\ 
\hline
April 05-07, 2000   & -292 & 01:00 & 098(07)  \\
August 11-13, 2000  & -234 & 10:00 & 225(12) \\
October 28-30, 2000 & -126 & 04:00 & 303(29) \\
March 19-21, 2001   & -149 & 14:00 & 079(20) \\
October 02-04, 2001 & -166 & 15:00 & 276(03)  \\
April 17-19, 2002   & -127 & 08:00 & 108(18) \\
June 17-19, 2003    & -141 & 10:00 & 169(18) \\
August 17-19, 2003  & -148 & 16:00 & 230(18) \\
July 24-26, 2004    & -136 & 17:00 & 205(25) \\
August 29-31, 2004  & -129 & 23:00 & 243(30) \\
May 29-31, 2005     & -113 & 14:00 & 150(30) \\
\hline
\end{tabular}  
\end{center}
\caption{Minimum Dst values with the corresponding Day of Year and Date (DOY(DD)) of minimum for the severe storms analyzed during 2000-2005.}
\end{table*}  

\newpage
\subsection{The severe storm of April 06, 2000}

On April 04, 2000, a CME event took place near the Sun's western limb. The CME shock hit the magnetosphere of Earth on April 6, 2000 \cite{sc:4}. This storm has been studied by several authors \cite{sc:32,sc:33,sc:34} and references therein. The event not only just resulted in the PPEF but also the DDEF and the travelling ionospheric disturbances \cite{sc:28} and references therein.
Figure 1 shows the variation of storm parameters during April 05-07, 2000. In Figure 1a, the SYM-H index has been plotted wherein the storm commenced at 16:45 UT on April 6, 2000. The SYM-H dropped to a minimum with a value of -320 nT at 00:09 UT on April 07, 2000. In Figures 1b and 1c, the variations of IMF $B_z$ and $B_y$, respectively, are shown. The IMF $B_z$ turned southward reaching below a value of -10 nT at 17:46 UT on April 06, 2000, reached a minimum of -33.0 nT at 23:12 UT and remained below -10 nT for a duration of 06:34 until 00:21 UT on April 07, 2000, thus indicating the storm to be of severe (G4) level according to the NOAA scales. During the period when IMF $B_z$ turned southward, was minimum and turned northwards, the corresponding IMF $B_y$ indicated values of -16.0 nT, 0.4 nT and -15.0 nT respectively.    
Furthermore, the variation of IMF B has been shown in Figure 1d wherein the value of B had been 19.1 nT during $B_z$ turning southward, 33.0 nT when $B_z$ dropped to a minimum value and 19.6 nT at $B_z$ turning northward.
Figure 1e shows the variation of the IMF clock angle, $\theta$ in degrees. The clock angle had a sharp transition from northward with a value of 14.115$^\circ$ at 17:48 UT to southward with a value of -1.168$^\circ$ at 17:49 UT on April 6, 2000.
\begin{figure*} 
\includegraphics[width=5.5in,height=4in]{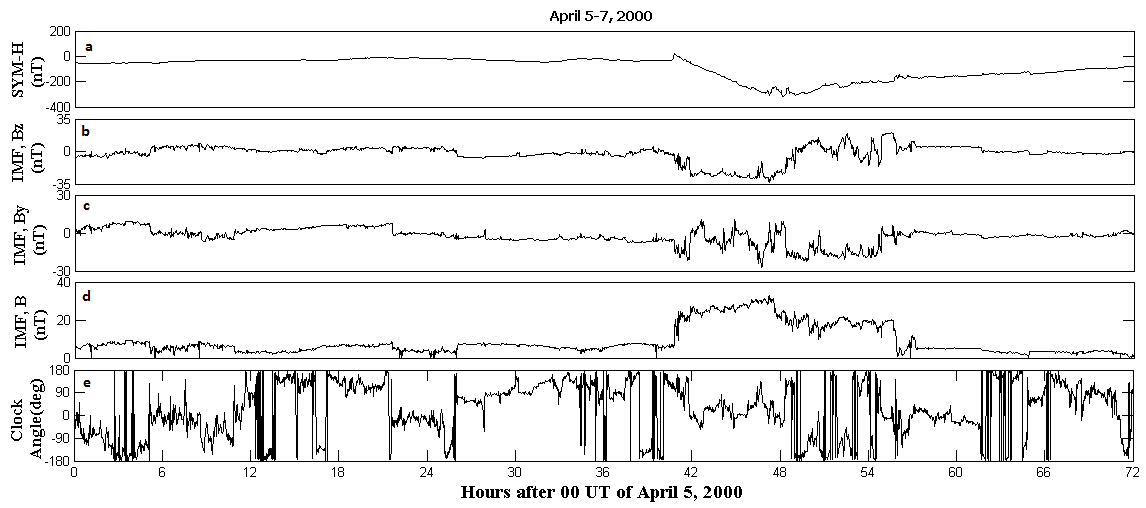}
\caption{Variation of interplanetary parameters during April 05-07, 2000. Panels a to e show SYM-H followed by IMF $B_z$, $B_y$, B and the IMF clock angle respectively.}
\end{figure*}

The global effect of the enhanced PPEF at dusk on the equatorial ionosphere has been observed by analyzing the in situ ion density measurements from successive DMSP transits crossing the equator between 19:00 and 22:00 MLT over the magnetic latitudes -30$^{\circ}$ to 30$^{\circ}$. Figure 2 show the plots of the total ion density for the orbits of DMSP f12, f14, and f15 satellites with equator crossing times ranging from 17.21 UT to 23.36 UT with corresponding MLTs ranging from 21.24 to 20.95 on April 6, 2000. 
Sudden outbursts of irregularities about the magnetic equator (indicated by black colored down arrows in this figure) can be observed around 35.35$^{\circ}$E at 18.30 UT (20.66 LT), 34.43$^{\circ}$E at 18.92 UT (21.21 LT), 09.40$^{\circ}$E at 20.01 UT (20.64 LT) and 355.52$^{\circ}$E at 20.67 UT (19.04 LT). 
Comparing the time of irregularity occurrence with the time of IMF $B_z$ crossing -10 nT during the main phase of the storm, it is found that the irregularity occurred with a delay of 3.23h, Furthermore, the irregularity occurred with a delay of 0.49h from the IMF clock angle transition from northward to southward.    
\begin{figure*} 
\includegraphics[width=\columnwidth,height=2.2in]{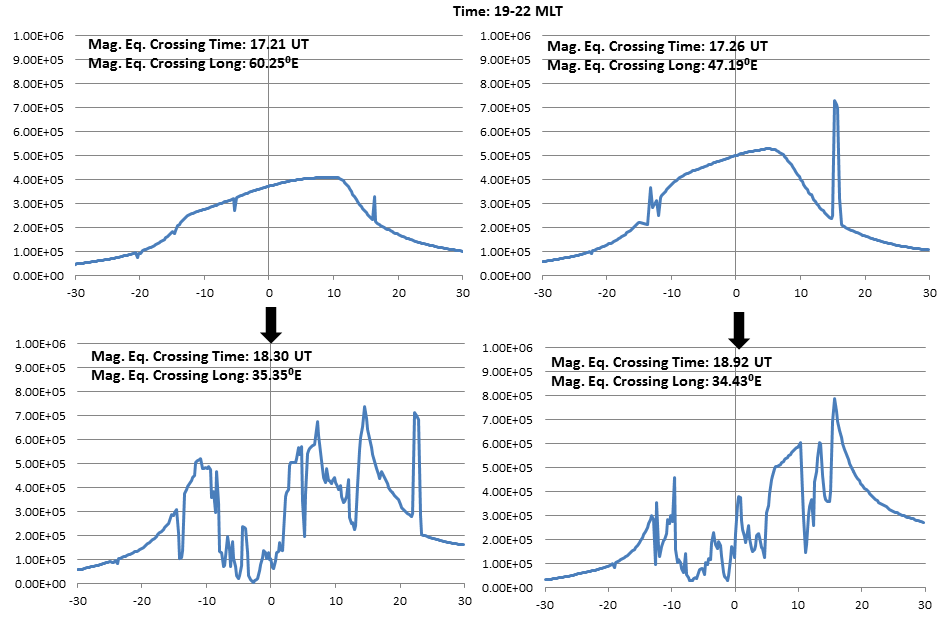}\\
\includegraphics[width=\columnwidth,height=2.2in]{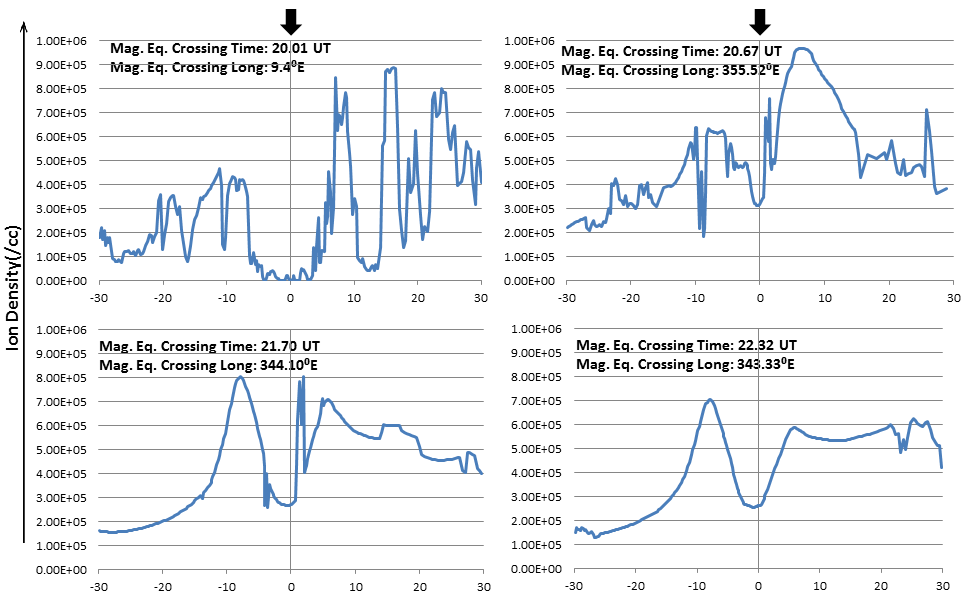}\\
\includegraphics[width=\columnwidth,height=2.2in]{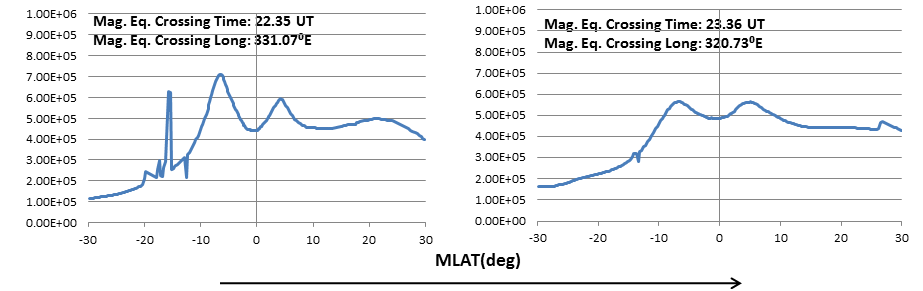}
\caption{Total Ion Density (cm${^-3}$) variation as observed by the DMSP f12,f14 and f15 satellites on April 06, 2000 from post-sunset to pre-midnight UT. The black down arrows indicate presence of irregularities during this period.}
\end{figure*}

\clearpage 
\subsection{The strong storm of May 30, 2005}

A CME hit the Earth's magnetosphere on May 29, 2005 at around 09:30 UT and caused a strong geomagnetic storm on May 30, 2005 (www.spaceweather.com). Figure 3, similar to Figure 1, shows the variation of different interplanetary parameters during the storm period of May 29-31, 2005.
In Figure 3a, the SYM-H index shows the storm commencement at 21:40 UT on May 29, 2005 with the minimum value dropping to -127 nT at 19:30 UT on May 30, 2005, signifying this storm to be strong. 
The IMF $B_z$ turned southward, reaching below the value of -10 nT at 04:27 UT on May 30, 2005 in Figure 3b while IMF $B_y$ showed -16.47 nT at that instant as observed in Figure 3c. IMF $B_z$ remained below -10 nT from 06:35 UT, with a value of -14.5 nT, to 15:34 UT with a value of -14.7 nT, on May 30, 2005, which is for a duration of 08:59. During this period IMF $B_y$ recorded -5.9 nT and -3.5 nT respectively.
From Figure 3d, the value of B was 19.7 nT when $B_z$ first turned southward and remained below -10 nT while showing a value of 15.2 nT at $B_z$ turning northward. Figure 3e shows the variation of the IMF clock angle. The clock angle had a sharp transition from northward with a value of 179.868$^\circ$ at 03:56 UT to southward with a value of -179.313$^\circ$ at 03:57 UT on May 30, 2005.
\begin{figure*} 
\includegraphics[width=5.5in,height=4in]{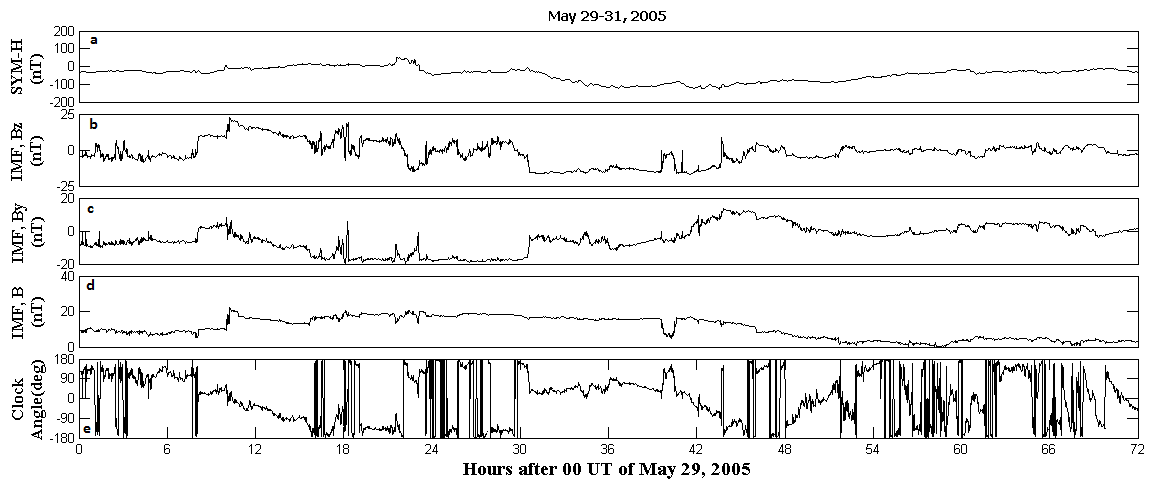}
\caption{Variation of interplanetary parameters during May 29-31, 2005. panels a to e show SYM-H followed by IMF $B_z$, $B_y$, B and the IMF clock angle respectively.}
\end{figure*}

Figure 4 shows the variations in the total ion density for the orbits of the three DMSP satellites with equator crossing times ranging from 09.26 UT to 17.80 UT on May 30, 2005. The only outburst of irregularity about the magnetic equator (indicated by black colored down arrow in the figure) is observed around 172.59$^{\circ}$E at 07.30 UT (18.81 LT) with the corresponding MLT 20.55. Comparing the time of irregularity occurrence with the time of IMF $B_z$, it is found that the irregularity occurred with a delay of 2.85h. Furthermore, the irregularity occurred with a delay of 3.37h from the IMF clock angle transition from northward to southward.     
\begin{figure*} 
\includegraphics[width=\columnwidth,height=2.2in]{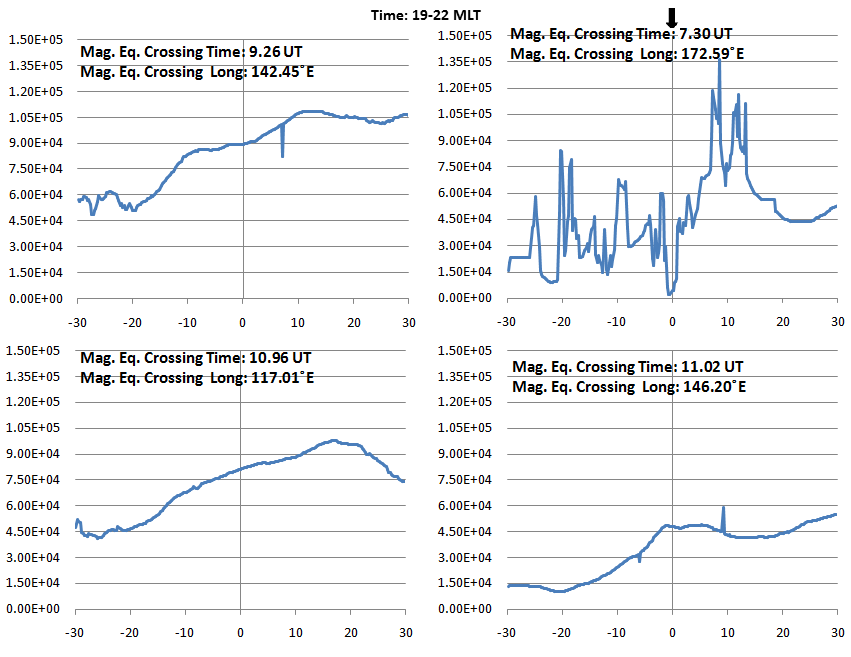}\\
\includegraphics[width=\columnwidth,height=2.2in]{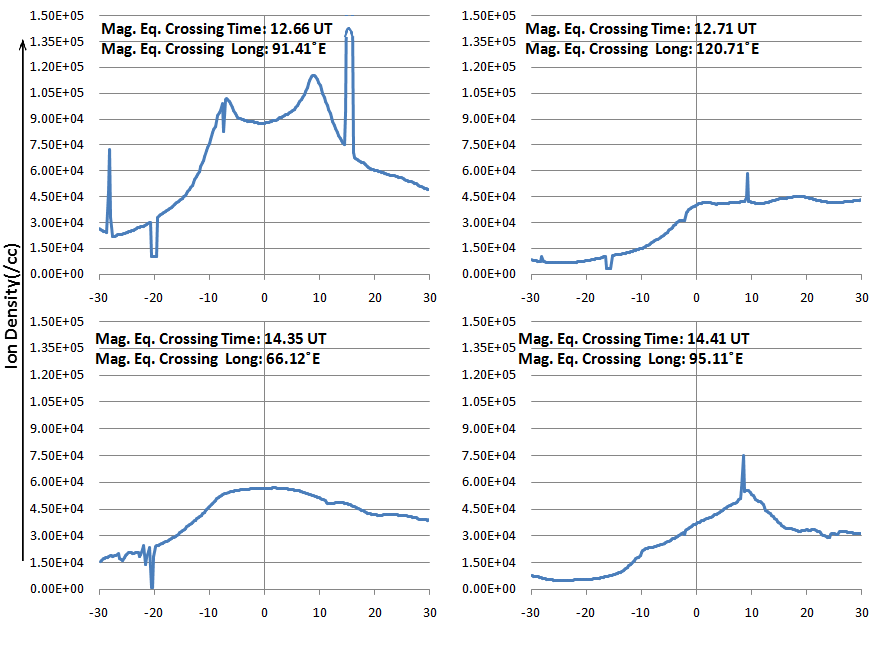}\\
\includegraphics[width=\columnwidth,height=2.2in]{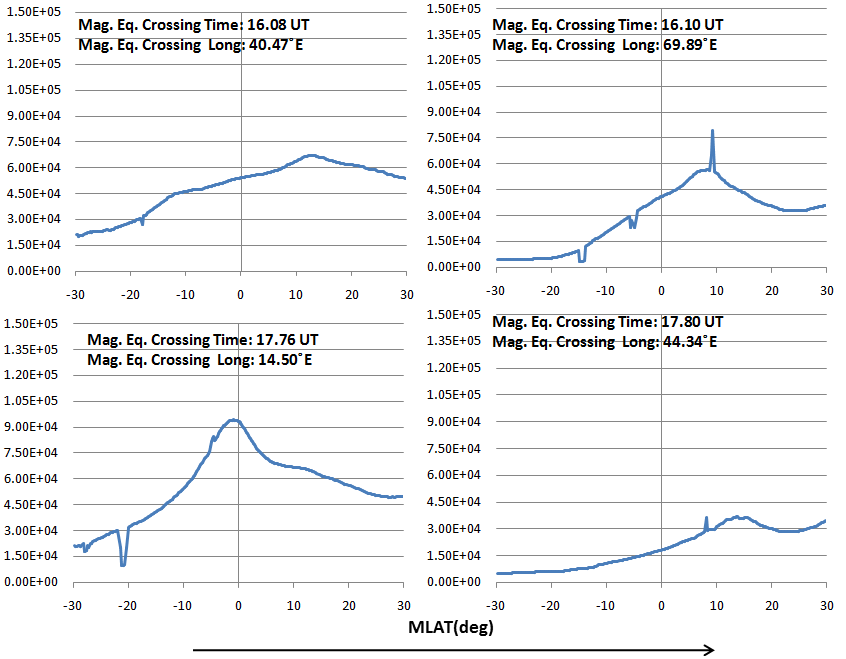}
\caption{Total Ion Density (cm${^-3}$) variation as observed by the DMSP f12,f14 and f15 satellites on May 30, 2005 from morning to post-sunset UT. Presence of irregularity is indicated by black down arrow.}
\end{figure*} 

Figure 5a shows the distribution of the time of transition of the IMF clock angle (red vertical bar) and the time of IMF $B_z$ crossing -10 nT (blue vertical bar) for all the eleven storms. Figure 5b shows the distribution of the delays of irregularity occurrence from the IMF clock angle transition (red vertical bar) and IMF $B_z$ crossing -10 nT (blue vertical bar).
The delay between irregularity occurrence from the time of IMF $B_z$ crossing -10 nT is below 4h for all the storms, which is in accordance with that reported in \cite{sc:7}. Additionally, it is observed that the delay between the IMF clock angle transition and the irregularity occurrence is below 3.5h for all the cases, which is an improvement over that reported by \citeA{sc:7}. Furthermore, for 91$\%$ of the cases, irregularity occurred within 3.5h and 3h from the time of IMF $B_z$ crossing -10 nT and the time of northward to southward transition of the IMF clock angle, respectively.
\begin{figure*}
\centering\includegraphics[width=\columnwidth,height=3in]{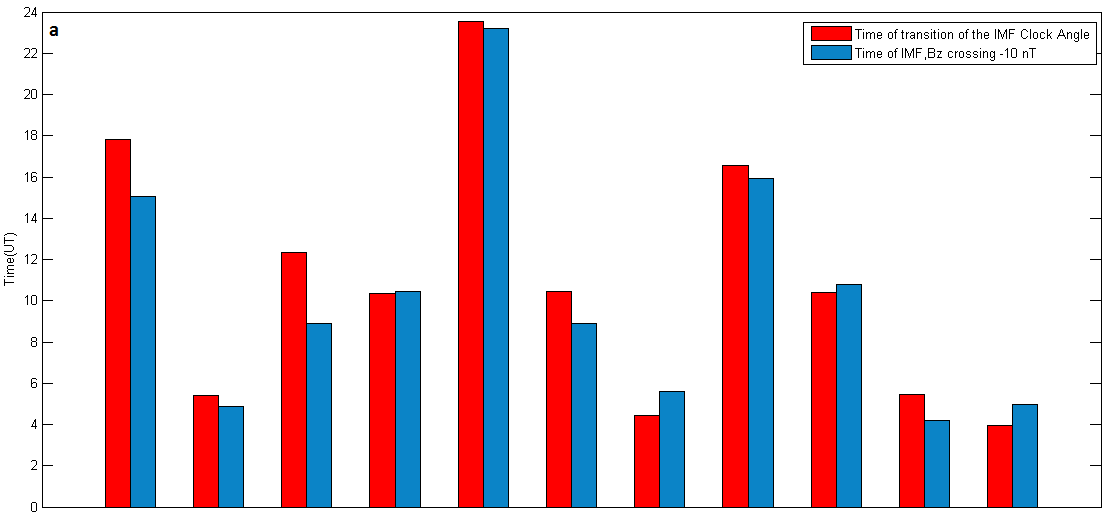}
\centering\includegraphics[width=\columnwidth,height=3in]{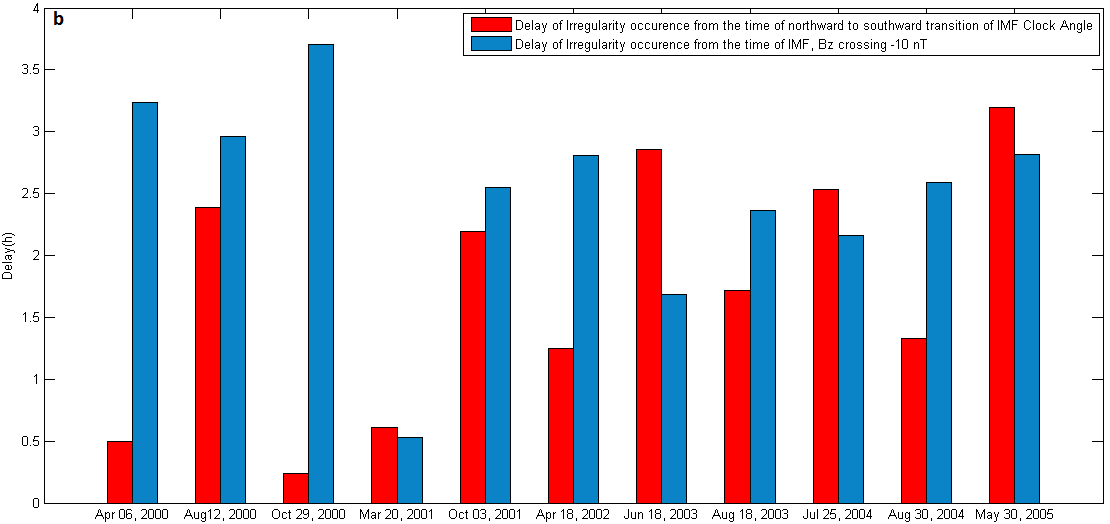}
\caption{(a) Distribution of the time of transition of the IMF clock angle (in red) and time of IMF $B_z$ crossing -10 nT (in blue); (b) Distribution of the delay of irregularity occurrence from the time of northward to southward transition of the IMF clock angle (in red) and the delay of irregularity occurrence from the time of IMF $B_z$ crossing -10 nT (in blue) for all the storms.}
\end{figure*}

Figure 6 shows the irregularity occurrence geographic latitude and longitude overlaid on the world maps (www.mcrenox.com.ar) for the all the storms. The rectangles in the figures designate the delay (given as colored bar ranging from 0h to 4h in steps of 0.5h) between the time of northward to southward transition of the IMF clock angle and irregularity occurrence longitude (the cyan cross marks) for the storms during 2000-2005.
It is clear from this figure that the irregularities for 10 out of 11 storms occurred within 3h from the time of northward to southward transition of the IMF clock angle and at a longitude sector having post-sunset local time. 
\begin{figure*}
\includegraphics[width=1\columnwidth,height=4in]{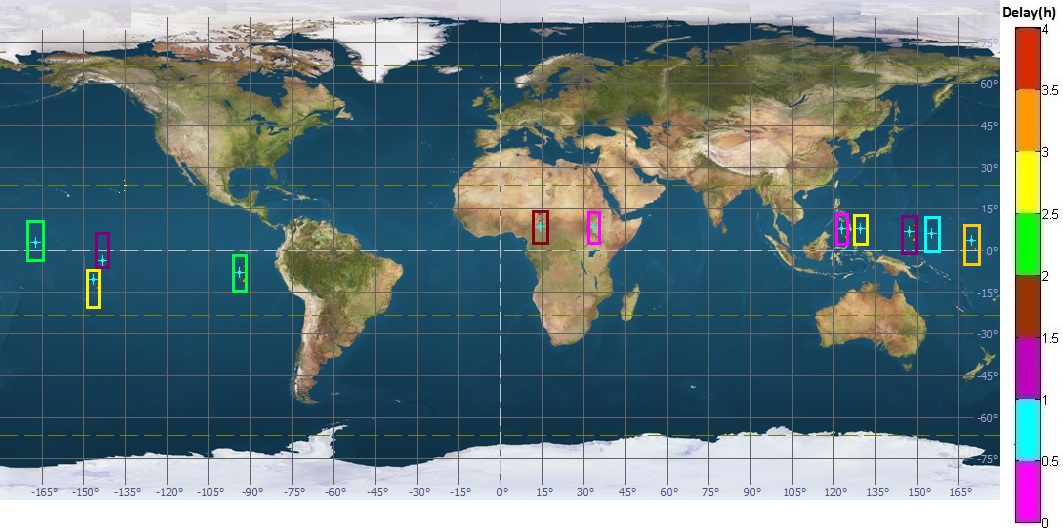}
\caption{World map showing the geographic latitude and longitude of the irregularity occurrence. The delay (in hours and designated as rectangles) between the time of northward to southward transition of the IMF clock angle and irregularity occurrence longitude (the cyan cross marks) for all the storms during 2000-2005 are shown}.
\end{figure*}

\clearpage
\section{Conclusions}

Study of the equatorial and low-latitude ionosphere, especially during geomagnetic storm time conditions, is useful to understand the dynamics and variability it presents, that would affect the navigation by satellite systems. In this paper, for eleven strong-to-severe geomagnetic storms during the declining phase of solar cycle 23 (2000-2005), it has been observed that for $\sim91\%$ of the cases, post-sunset equatorial irregularities occurred within 3.5h from the time of IMF $B_z$ crossing -10 nT and within 3h from the time of northward to southward transition of the IMF clock angle.  
\citeA{sc:7} reported that within 4h of the southward IMF $B_z$ crossing -10 nT, irregularity would occur in the dusk longitude sector.
For predicting the storm time occurrence of ESF in response to PPEF, when undershielding condition prevails, the clock angle transition time provides better accuracy than the time of $B_z$ crossing -10 nT, as evident from this study based on eleven strong-to-severe geomagnetic storms. 
This study also shows the importance of taking the $B_y$ component of IMF into account, in addition to the $B_z$ component as IMF $B_y$ plays an important role in determining the PPEF polarity \cite{sc:25,sc:26}.
This paper, for the first time, shows that by having the knowledge of the time of sharp transition of the IMF clock angle, it would be possible to predict the longitude sector that would be affected due to the ESF generation, thus an improvement, and hence a better forecast lead time, from the previously reported 4h window of ESF generation from the southward IMF $B_z$ crossing -10 nT.

\acknowledgments
SC acknowledges Space Applications Centre (SAC), ISRO for providing fellowship under project NGP-17. The authors acknowledge the Center for Space Physics, University of Texas at Dallas and the U.S. Air Force for the DMSP plasma data. Acknowledgements go to NASA Goddard Space Flight Center-Space Physics Data Facility (GSFC-SPDF) for the 1 minute high resolution omniweb data available at https://omniweb.gsfc.nasa.gov
/form/omni$\_$min.html for the SYM-H index and the interplanetary parameters: IMF $B_y$ and $B_z$. Further acknowledgements go to the World Data Center (WDC) at Kyoto University for the Dst index data available at http://wdc.kugi.kyoto-u.ac.jp/. SR would like to thank International Centre for Theoretical Physics (ICTP), Trieste for providing support through Senior Associateship Program. 

\clearpage
\bibliography{agusample}

\end{document}